# Exploring and Benchmarking High Performance & Scientific Computing using R, R's HPC Packages and Lower level compiled languages : A Comparative Study

Rahim K. Charania
The Pennsylvania State University, University Park, Pa
Graduate Research Assistant – Institute of Cyber Science.
rkc10@psu.edu

*Abstract*—R is a robust open-source programming language mainly used for statistical computing . Many areas of statistical research are experiencing rapid growth in the size of data sets. Methodological advances drive increased use of simulations. A common approach is to use parallel/concurrent computing. This paper presents an overview of techniques for parallel computing with R on ACI (a PSU Infrastructure) and benchmark it with C/C++. We review the scalabilty concern of R, and look at the simplicity of using R as a primary language in Coding for HPC. We will look at the various R packages for HPC like Rmpi, Rcpp, snow and snowfall. We utilize a series of algorithms to benchmark and will illustrate each benchmark with a representative graph for ease of understanding. The paper concludes with a better understanding of which language to use when in high performance computing .

## I. INTRODUCTION

With over 10000 libraries already available in its central repository, CRAN, covering state-of-the-art statistics, data-mining and machine-learning libraries almost anyone with a Scientific aptitude can hit the ground running with R. With this versatile language you often don't need to write code at all. R , an Open Source Programming language, for statistical computing, is highly extensible through the use of packages. These packages are hosted at the CRAN Repo(Comprehensive R Archive Network) as stated initially. "Packages" are libraries for specific functions or specific areas of study, frequently created by R users and distributed under suitable licenses. The R Syntax is designed for analysis of data & Rapid prototyping. Hence, the recent major interest in people from the Statisticians and Bio-Engineers and the likes. Providing software for parallel or high performance computing (HPC) with R was not a development goal. However, there is growing interest in using it for HPC. The 2 primary drivers for the increased focus can be attributed to the (a.) Growing data sets (Immense amount of data being collected, case in point: Facebook),& (b.) Increased computational requirements that stem from complex algorithms to analyze and use the data.

Now turning our attention to the real question: with the abundance of coding languages and scripting languages for interpretation, why do I use R? Well, to answer that question let me refer you to the varied Fortune 50 companies including but not limited to Google, Facebook, Twitter, Microsoft, IBM, and the list goes on, use it for Exploratory Data Analysis, Experimental Analysis, Big Data Visualization, Anomaly Detection and a million other uses ranging from Brain Scan MRI imaging analysis to understanding user behavior, right upto the Wall Street firms that use it for Stock Prediction and Portfolio optimization, R can pretty much do everything. IEEE Spectrum ranked it 6$^{th}$ in it's yearly review of top Programming

*Illustration 1: IEEE Spectrum 2017 Ranking*

Languages, ranked just below C# and its close sister Python, which tends to be more versatile in the coding perspective, but has a steeper learning curve and is not as "user-friendly" compared to R.

## II. EASE OF USE & RELATED WORK

### A. Ease:

One can rapidly prototype in the R language. R is an interpreted language; it will dynamically convert types. It's also an excellent tool for visualization and analysis (example; the diverse GGplot2 library). There is also a huge community of R developers, creating new solutions for problems continuously.

Moving from C/C++ to R not only increases productivity but also shrinks code line-counts for similar tasks by a sizable amount as we will see in the further sections. To give a small flavor of why, consider one of the most common patterns of iteration over some collection, selection and action:

---

"R is a shockingly dreadful language for an exceptionally useful data analysis environment." — Tim Smith

*declare iterator for collection*

   *for element in collection*

     *if (element meets condition)*

        *do something with element*

In R this construct typically shrinks to:

   *collection[condition] <- new_values*

Where new_values is a vector expression as well. And this is not only limited to arrays/vectors.

Also there is no need to declare iterator-variables and write loops to iterate over them because when acting on vectors (or higher dimensional data-structures) the loops are implied.(See Section V-D)

R has sophisticated debugging utilities for dealing with buggy R code. (eg. Valgrind, debug(), str(), etc.). Using gdb is possible, but honestly is unnecessarily complicated. Instead, using Valgrind is extremely straight-forward and pretty easy.

### B. Pitfalls of R and where C/C++ blooms:

R, compared to low-level languages such as C++, improves development time at the expense of runtime. Fortunately, there are a handful of ways to speed up operation runtime in R without sacrificing ease of use. Optimizations and tricks to increase utility in production and testing are discussed further and are the primary objective of this paper.

It is safe to say that R can be pretty slow at times, and it's memory-intensive. Importing the data sets is very easy and fast. It's the ballooning memory requirements for actually processing that data that's the problem. Even with all that, though, R is a very useful tool to have in your toolbox. But it's not exactly a computational powerhouse. C/C++ are insanely low productivity language, but, tend to be better than R for slow for iterative, procedural code.

### C. Related Works:

There is not a lot of literature on utilizing R for HPC however, the resources that are available are pretty robust. 2 names stand out in R's documentation for HPC Dirk [1] Eddelbuettel and Glenn K. Lockwood. These two Computational Scientists are at the forefront of innovation at using R for HPC. They have published books and several blogs about these topics which are detailed in the list of references and are very active on R mailing lists and discussing issues with coders. There is some benchmarking done for R packages but it is very trivial and not a lot of people have looked into benchmakring R with lower level coding languages like done in this paper and using scalability to check each parallel packages limits and uses. This paper would behave more as a beginners guide to HPC using R and scientific computing. We will start with the most famous and widely used packages and lead up to a Neural Network.

### III. SYSTEM MODEL

We start off with demonstrating why to use R which we have done above and then we move on to it's practical applications like using it in a neural network and benchmarking it against Lower Level Languages, and we also see various scalability models for the various packages we have and the advantages of using R over the latter. A small snippet of the code using R is provided (since R is pretty nifty with smaller lines of code), however, for the C/C++ aspect the entire code can be found in the zip file attached and only pseudo code or major steps are shown here if at all due to space constraints. After each codes results there is a small section detailing the analysis of why the system acted the way it did and how and why we have received these answers, and these timings from the code. Complexity of the code through each passage is gradually increased for ease of user understanding (since our primary audience will be R novices).

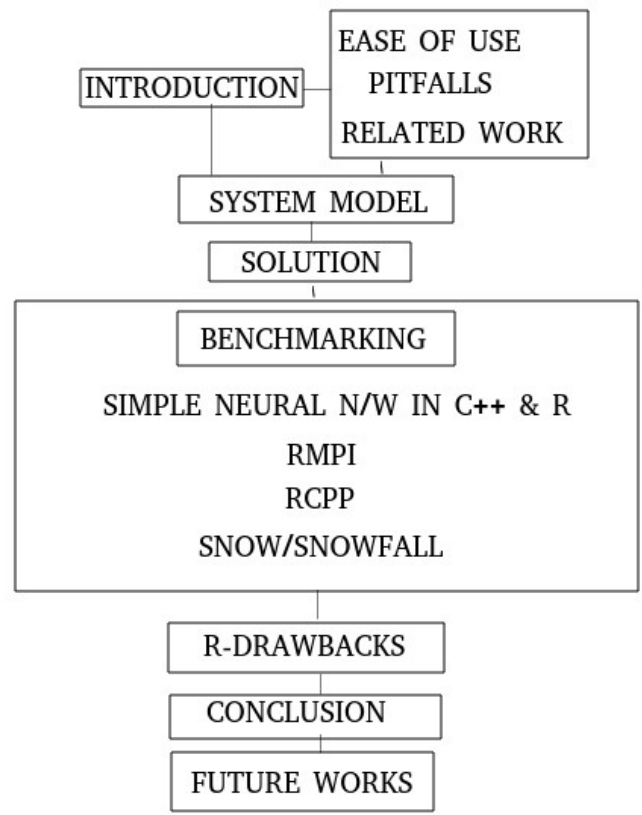

*Illustration 2: System Model*

### IV. SOLUTION

In the next section we see benchmarks and code analysis of R, it's various packages and C/C++ and scalabilty of R. From the findings it is clearly evident that R is pretty easily typed

The greatest weakness and greatest strength of R is that it is not a strongly typed language. Therefore easy tasks in strongly typed languages such as re-factoring, auto-compiler checks, unit testing, etc. can be more difficult in R.

C++ is not necessarily "faster than R". It makes limited sense to talk about faster languages, only faster implementations. Yes, it is more robust and easier to tweak at the ground level as seen further but it is important to understand that not everyone requires such minute and precise control over their code intricacies.

An example would be garbage collection and memory allocation. A biologist sequencing DNA and carrying out EDA of the genome does not need to know malloc() and free() constructs but can just freely call a vector and assign it a huge value with 1 line of code. However, this tends to be extremely slow for larger arrays in R as opposed to C but the benefits like lower code complexity and easier syntax, outweigh the few seconds lost in reading an array.

C++ overcomes the above issues and gives you more room for optimization by giving you deterministic control over low level facilities that allow you to manually manage memory and access patterns. That's all. Chances are, a naive C++ implementation of native R routines will be slower than using the R implementation.

The following table is a pretty good representation of the comparison between R and C/C++, where we can clearly see the specific areas they excel in:

| Language/Attributes | R | C/C++ |
|---|---|---|
| **Performance-wise** | | ✓ |
| **Ready Packages available** | ✓ | |
| **Ease of EDA (Exploratory Analysis)** | ✓ | |
| **adding GUI like features and interactivity** | | ✓ |
| **Coding Small Projects** | ✓ | |
| **Coding large projects** | | ✓ |
| **Scalability** | ✓ | |
| **Ease of learning eg. syntax** | ✓ | |

Having seen these now we delve into actually see what's going on under the hood and get our hands dirty with coding, we will also see how to overcome the C++ advantages where R fails to excel by integrating C++ seamlessly into your R code, too.

## V. BENCHMARKING

Now we see benchmarking :

### A. Simple Primer to Scientific Computation, Neural Network:

To start off understanding benchmarks for R HPC packages against C/C++ we need to have a base line of the understandings between R and C/C++. Let's see a case where R excels, Statistical Computing. We will now see a neural network in both R and C. We use IRIS datasey which is an open source flower dataset which predicts the flower genus by different dimensions and attributes. The dataset is available on the UCI repository.

The R code is super-compact, but the C++ code is extremely lengthy and cumbersome. This attribute of R shows how nifty it is. Lets go ahead and see the R code and benchmark it against C++.

| Languages | C++ | R | Speedup |
|---|---|---|---|
| **Time(sec)** | 0.0338 | 15.10495 | 446.89201183 |

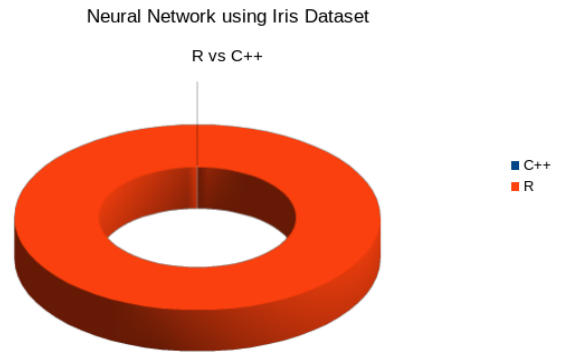

Neural Network using Iris Dataset
R vs C++

In terms of speed and performance C++ blew R away by a long shot. However the key problem with C++ was the complicated syntax and the lengthy program. Also we know it is not best suited for Stats and Machine learning for the ease of coding. We need to code everything from scratch. Getting the data is also a cumbersome process as opposed to R where you can call the entire csv file with 1 command i.e. read.csv("file"). The code for R is 10 lines and C++ is over 100's of lines.

Now coming on to the nitty gritty of the code and algorithm, the C++ code is ~4000 times faster at execution as opposed to the R code. But given it's a serial code we can understand and we will be seeing parallel speedups in the further sections which will not be as fast as this serial code.

*R code main part:*

*iris_net <- neuralnet(f, data = pre_process_iris, hidden = c(16, 12), act.fct = "tanh", linear.output = FALSE)*

The Time complexity of a Multi Perceptron Neural Network is $O(2n)$ to converge to the optimal solution, this is because $O(no.(epochs)*no.(egs.)*no.(features)*no.(neurons))$, this leads to the approximate solution.

This example was just to get started with the core concepts of C++ and R in Scientific computing now we will see parallel packages to exploit parallelism and concurrency in code.

Let's now take a look at R's 3 to 4 most popular and versatile parallel packages used widely in academia and industry.

### B. Rmpi:

Rmpi is an interface, or wrapper, to MPI. It provides an interface to low-level MPI functions from R so that users do not have to know details of the MPI implementations. It provides an interface necessary to use MPI for parallel computing using R. As in MPI, Rmpi has master threads which fork/spawn into slaves to complete some tasks.

Master → Main CPU

Slaves → dependent CPU's

Command to spawn a slave: *mpi.spawn.Rslaves()*

Using this command first gets number of CPU's available. Now we can use "nslaves" option to specify a specific number of CPU's.

Specifying a higher number of CPU's than available on the system will not return an error. But instead will just work with as many CPU's are available on the System.

The package provides several R-specific functions, in addition to wrapping the MPI API. For example, parallel versions of the R apply-like functions are provide by, e.g., mpi.parApply(). Also, R objects can be efficiently sent to slaves using mpi.bcast.Robj2slave().

Like send and receive in the primitive MPI in Rmpi we can use function calls like mpi.remote.exec() and mpi.bcast.cmd() to communicate with the slaves. (Fun Fact: In Los Angeles, officials pointed out that such terms as "master" and "slave" are unacceptable and offensive, and equipment manufacturers were asked not to use these terms. )

Bcast command works exactly like the available broadcast call in MPI and the slaves will execute the commands but there will be no return form them. Remote.exec() gives return form slave.

Now we look at a simple prefix sum using the rank of slaves as our input:

> mpi.remote.exec(sum(1:mpi.comm.rank()))

X1 X2 X3 X4 X5 X6 X7 X8 X9 X10

1  1  3  6  10  15  21  28  36  45  55

Although it doesn't have all commands found in original MPI for C/Fortran, quite a few functions have been added and it has most of basic functions for normal operations.

Now that we have our basics for Rmpi straight and have seen a trivial example lets go ahead and code a program that averages 1 million random number s in C and R using MPI and Rmpi, respectively.

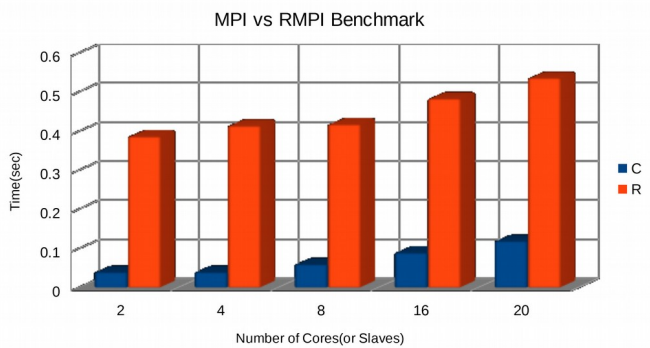

| Slaves | C | R | Speed-Up of C over R |
|---|---|---|---|
| 2 | 0.04 | 0.3873832 | 9.68458 |
| 4 | 0.04 | 0.4151559 | 10.3788975 |
| 8 | 0.06 | 0.4185169 | 6.9752816667 |
| 16 | 0.09 | 0.4844346 | 5.3826066667 |
| 20 | 0.12 | 0.5371654 | 4.4763783333 |

From the Output we can clearly see that C has a clear advantage in terms of speed of execution for a large array arithmetic. Using C and MPI we can achieve a 500-1000% performance increase. However, it is key to note that the R script was only 4 lines as opposed to the C code which was almost 75 lines. The algorithmic complexity was obviously the same, however, it is also important to note the trend form the graph that R scales better than C since with 2 cores/slaves we observed a 1000% increase and at 20 cores the C performance boost almost halved to 500%.

Now since we know R scales well with Rmpi, lets specifically see a case of Rmpi with multiple iterations of the same averaging case. We iterate the average of 1 million, 400 times

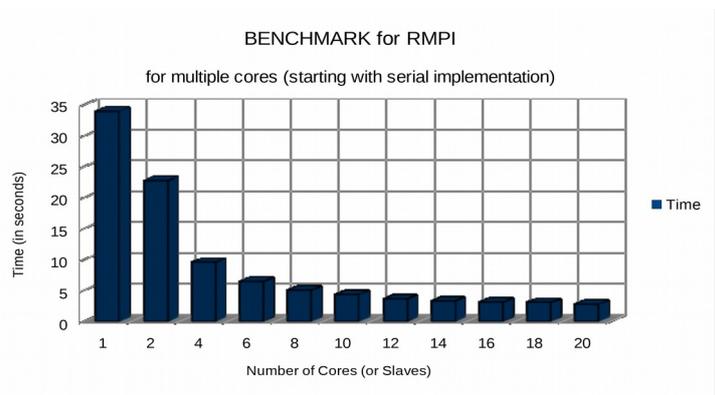

and benchmark to see if we can find a pattern in scalability, the results were as follows:

RMPI Benchmark Output:

**Serial: 1 core**     33.98278 sec

| Slaves | Time (secs) | Speedup |
|---|---|---|
| 2 | 22.81022 | 1.48980500845674 |
| 4 | 9.520365 | 3.56948289272523 |
| 6 | 6.536435 | 5.19897773021532 |
| 8 | 5.073261 | 6.69840956339522 |
| 10 | 4.327422 | 7.8528925535804 |
| 12 | 3.597253 | 9.44686959744005 |
| 14 | 3.258143 | 10.4301069658391 |
| 16 | 3.130024 | 10.8570349620322 |
| 18 | 3.048821 | 11.1462037292448 |
| 20 | 2.784601 | 12.203823815333 |

In the example for Rmpi scaling as shown above we only has to change 1 value from the prior code that is all, no lengthy declarations or functions:

*Code snippet:*

*> library('Rmpi')*

*> mpi.spawn.Rslaves(nslaves=20)*

*>b=Sys.time()*

*>mpi.iparReplicate(400,mean(rnorm(1000000)))*

*> Sys.time()-b*

Hence, as seen above the power lies in its simplistic design and ease of use. The code size is cut down to less than 10% as opposed to lower level languages. Syntax is similar and self explanatory. We can see that R scales fabulously with increasing the number of cores in Rmpi and hen it plateaus off after 20 cores, upto 24 cores were used in this series of testing.

The speedup achieved is phenomenal form the serial implementation costing ~34 seconds, and the fastest 20 core implementation costing only a mere 2.78 seconds. We achieved a speedup of ~1220% from a serial to a 20 core implementation.

## C. Rcpp:

> "It's compiled, not just scripted" Well, partially!

Sometimes R code just isn't fast enough. Profiling, debugging, optimizing all said and done, you still have that lingering feeling, could I have written this one function in C++ and made this code super fast. Well now you definitely can! Rcpp one of the most used and a very versatile package is a god send for coders who love low level coding and want to optimize and make their code even faster and closer to the compiler as ever. This tool was written by Dirk Eddelbuettel and Romain Francois, the R for HPC power users.

Rcpp makes it very simple to connect C++ to R. It provides a clean, approachable API that lets you write high-performance code. Now comes the question why should you use this? Well, in my experience in using R for Statistical Analysis, I have noticed these typical bottlenecks in R that can be easy to overcome with C++:

1. Some loops can't be easily vectorized since some of further iterations are dependent on prior ones.

2. Recursive Functions: Let's say we need to call a function a million times. Although not impossible in R, it has a large overhead. As opposed to C++ where it is much lower. Hence in such cases C++ seems to be a better choice.

3. Data Structures and Algorithms: Some code requires Advanced Data Structures and Complex Algorithms, which are not really practical in R.

To give you a basic overview of how to Rcpp syntax works, lets see a simple addition code using C++ embedded in R:

*> library(Rcpp)*

*> cppFunction('int add(int x, int y, int z) {*

*+ int sum = x + y + z;*

*+ return sum;*

*+ }')*

*> add(1,2,3)*

Answer=6

Of course there are more complicated code with classes that cannot be accommodated using this syntax for that we have sourceRcpp(), where we source the C++ code in R using the above mentioned call.

The same C++ code that is used with sourceCpp() can also be bundled into an R package. There are several benefits of moving code from a stand-alone C++ source file to a package:

1. Your code can be made available to users without C++ development tools.(viz. Github or maybe CRAN if approved)

2. Multiple source files and their dependencies are handled automatically by the R package build system.

3. Packages provide additional infrastructure for testing, documentation, and consistency.

Now, we will see one more complicated and sophisticated problem demonstrated and then benchmark it in both R and C++ and then see its ease of implementation in R and C++ using the Rcpp construct.

Let's dive into Stats and see a VAR, a Vector Auto regressive function. It basically has k number of endogenous variables $x_t$. A VAR(p) process is defined by a series of coefficient matrices $A_j$ with $j \in 1, \ldots p$ such that:

$$x_t = A_1 x_{t-1} + \ldots + A_p x_{t-p} + u_t$$

Now usually when we study VAR systems we leverage simulation to access these models

. Now for our benchmark we are going to see 2 examples one VAR(1) process of order 2 generation written in R and another in a C++ function to generate the VAR Simulated data. We are going to benchmark these 2 processes to check which one works faster in our case.

*#a=parameter and u=error, generating matrices*

*a <- matrix(c(0.5,0.1,0.1,0.5),nrow=2)*

*u <- matrix(rnorm(10000),ncol=2)*

*R solution:*

*> rSim <- function(coeff, errors) {*

*simdata <- matrix(0, nrow(errors), ncol(errors))*

*for (row in 2:nrow(errors)) {*

*simdata[row,] = coeff %*% simdata[(row-1),] + errors[row,]*

*}*

*return(simdata)*

*}*

*> rData <- rSim(a, u)*

*This approach is simple* and straight forward. The simulation function receives a 2 by 2 matrix of parameters and N by 2 matrix of errors i.e. randomly distributed.

It then creates a result vector of dimension N by 2.

*Now we takea look at the C++ solution:*

*suppressMessages(require(inline))*

*code <- 'arma::mat coeff = Rcpp::as<arma::mat>(a); arma::mat errors = Rcpp::as<arma::mat>(u); int m = errors.n_rows;int n = errors.n_cols;*

*arma::mat simdata(m,n); simdata.row(0) = arma::zeros<arma::mat>(1,n);*

*for (int row=1; row<m; row++) {simdata.row(row) = simdata.row(row-1)\*trans(coeff)*

*+ errors.row(row);} return Rcpp::wrap(simdata);'*

*rcppSim=cxxfunction(signature(a="numeric",u="numeric"),code,plugin="RcppArmadillo") rcppData <- rcppSim(a,u) # generated by C++ code*

*stopifnot(all.equal(rData, rcppData))*

Having run a comparison to check the time taken by both these approaches we see:

|  | **Time** |
|---|---|
| **rSim** | 0.03585243 |
| **rcppSim** | 8.215408 |

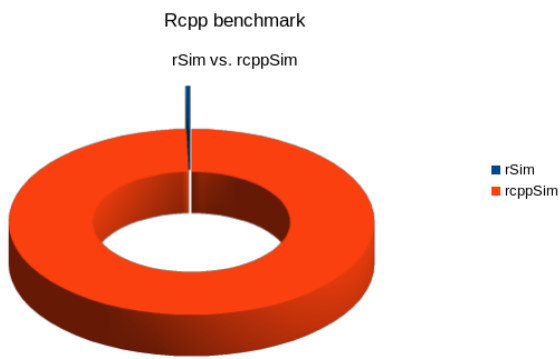

Now having run the benchmark it is clearly visible that the R Solution takes only a fraction of the time the C++ solution would take. It is approximately ~200 times faster than the C+ solution in this case.

Hence, we have now seen that code can be written in C++ which is very similar to the R code first used to prototype a solution. Thanks to tools such as the inline package and particularly the more recent Rcpp attributes, we can easily extend R with short C++ functions—and reap great performance boosts in the process.

*D. Snow:*

Snow (an acronym for Simple Network Of Workstations) provides a high-level interface for using a workstation cluster for parallel computations in R. It relies on the Master / Slave model of communication in which one device or process (known as the master) controls one or more other devices or processes (known as slaves). It's pretty widely used and was one of the earlier packages for parallelizing Code. Among the functions that this package provides, I found the parApply family the most useful ones and the easiest to work with to take profit of parallelism.

Snow works best for 'embarrassingly parallel codes,' eg, a simulation study, bootstrap, or a cross-validation. It is built on top of rmpi or sockets.

In order to use these functions, it is necessary to have firstly a solid knowledge of the apply-like functions in traditional R, i.e., lapply, sapply, vapply and apply.

If the use of apply functions is clear, then parallelisation is just one small step beyond with snow. The functions equivalents are:

| R | Snow - Package |
|---|---|
| lapply | parLapply |
| sapply | parSapply |
| vapply | – |
| apply(rowwise) | parRapply, parApply(,1) |
| apply(columnwise) | parCapply, parApply(,2) |

Now to understand this example we need to understand what apply actually does? Well it very simply just "applies" a function to an array/vector and gives a return.

Generally speaking we could see these Apply Functions As Alternatives To Loops.

- Bye, Bye For loops!

Now lets see a case where we replace a for loop with this apply-family function, let's begin with a trivial application of adding 2 to a array of integers from 1 to 100, We can see 2 implementations 1 in C++ and it's compact 1 line implementation in R using vapply:

**C++ pseudo code:**

*for (int i=0; i<1000000; i++)*

    *A[i]=i+1;*

*for (int i=0; i<1000000; i++)*

    *result[i] = A[i]+2;*

**R code:**

*A=1:1000000*

*result=vapply(A,function(x) x+2,FUN.VALUE=0)*

Both the above trivial examples provide the same output. The major difference lies in the simplicity

BENCHMARK:

| **BENCHMARK** | **C++** | **R** |
|---|---|---|
| Time | 2.29 | 0.5441196 |

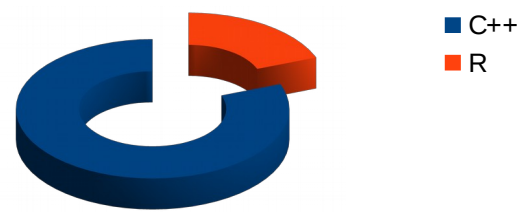

Again from this example we see that the R code is so much smaller than C++. Not only that but it is also faster in this trivial case. We also do not have the hassle of memory allocation and freeing, eg. malloc(), free() in R, simple call a variable as a vector and you are good to go!

Now that we have seen a side by side comparison of getting rid of for loops using vectors between R and C++, we go ahead and take care of exploiting parallelism with snow in R and benchmark it's timings with several nodes.

We move on to code a program which creates a random data frame with 3 attributes with the maximum element size of 1 million, and values ranging from 1 to 1 million, The R code is very simplistic as opposed to C/C++ which would have been several lines of code. We take those 3 elements and do a specific series of math operations to see which is the most optimal cost.

*Making Data Frame:*

> data <- data.frame (a = seq(1,1000000,1), b= seq(10,1000000,10), c=runif(100000))

*Serial code:*

> s <- apply(data,1, function(x) (x[1]*x[3]+x[2])*x[3]-9 > 20)

*Parallel Code:*

> Mycluster <- makeCluster(20) ## no. of cores in brackets

> parallel <- parRapply(Mycluster,data, function(x) (x[1]*x[3]+x[2])*x[3]-9 > 20)

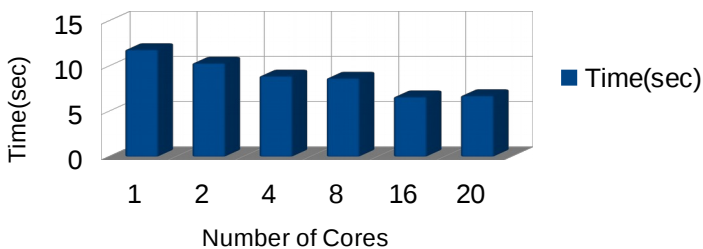

Benchmark Snow
Using different Cores

| Serial: 1 core | 11.82672 | |
|---|---|---|
| Slaves | Time (secs) | Speedup |
| 2 | 10.32954 | 1.1449415947 |
| 4 | 8.899717 | 1.3288871994 |
| 8 | 8.669181 | 1.3642257556 |
| 16 | 6.602432 | 1.7912672179 |
| 20 | 6.731653 | 1.7568820021 |

We again see perfect scaling with R, as we can see we achieve speed up ranging from ~1.44 to 1.79 when we increase the number of cores. We see apeak at 16 cores, and the speed reduces again after 16 when we go to 20 bumping up system time by ~0.13 seconds. SO we could sy it plateaus after 16 and then goes up as we increase the number of cores.

Hence, we again see from this example that scales very well.

*Snowfall:*

Snowfall is a top-level usability wrapper for snow to make parallel programming even more easy and comfortable.

Steps to use snowfall:

1. Detect which parts of code can be parallelized.

2. Change "for loops"→"apply-family" operators like lapply, vapply, etc.

3. Then write a wrapper function to be called by the list operators and manage a single parallel step.

It's pretty much standard and can be used exactly like snow. But now lets talk about an important concept in R that arises from this concept: variables and privacy. There are no local variables. Only the data from the list index will be given as argument. If you need more than one variable argument, you need to make the required variables global (assign to global environment) and export them to all slaves. Snowfall does provide some functions to make this process easier.

VI. DRAWBACKS OF USING R OVER LOWER LEVEL LANG.

All being said and done, is easy to write inefficient R code, both in terms of memory and speed. The most common cause as observed by various authors is building a data-frame iteratively , like adding one column at a time, using cbind() or similar, because this appends a column to the end of a data frame, and hence the copy-arguments on every call makes this $O(n).O(n)$ process become an $O(n^2).(n^2)$ process. Similarly, passing large data-frames as function arguments, when you only want to pass one or a subset of the columns, obviously has its (pass by value copy) undesired cost. If this is your cause for slow down, you want to look at data.table library which allows passing args by reference using :=, and learning about the <<- operator.

However, using Rcpp it is possible to avoid the above pitfalls where it matters and write the low level critical loop code in C/C++ where needed.

Memory Management: A critical aspect of programming predicts how much memory is needed and make the most of it. Efficient memory management enables us to write faster code and is a major cause of slow code. The building blocks of memory management are objects, functions, largerblocks of code. In R the base size of each object is 40B: metadata, 2 pointers to previous and next datapoint, one pointer to attributes, length and padding allocates vectors that are 8,16,32,64 or 128 bytes long. Beyond 128 bytes, R asks for more memory in multiples of 8 bytes. Shared objects are stored once and the global string pool is the 1 place where unique strings are stored.

R, unlike lower level languages, does a poor job of handling multiple different computations at the same time. The limitations of the R kernel are well-documented in Sridharan and Patel (2014): "R spends more than 85% time in processor and memory stalls High rate of cache misses (about 90% on linear regression and k-means tasks) Triggers garbage collection very frequently Creates a large number of unnecessary temporary objects, resorting to swap space quickly even for datasets that should have fit main memory"

## VII. CONCLUSION

One possible avenue we could explore is to use R for testing and C++ for implementation, given it's simplistic nature and ease of use, and C++'s robust concurrency handling.

Hence, in conclusion we can say that we successfully saw R has excellent scaling, it's great for Exploratory Data Analysis, has a ton of packages readily available which make coding form scratch a non-necessity, self explanatory commands/ syntax make it all the more easier and is perfect for small projects, however, for coding large projects C/C++ seem perfect unless you want to go into the nitty gritty and code the recursive loops in C/C++ to iron out the details using C++, and use RMPI for Multi node processing, it seems like that would create an ideal case scenario and is an excellent resource for data analysis and number crunching in today's big data phase where data and compute requirements are seeming to grow exponentially, we need something like R to utilize HPC and not worry about the low level details like memory allocation, and and repetitive function calls causing segmentation faults.

Side-note: Unless you are actually doing something like High Frequency Trading you do not need the true blistering performance boost a pure C++ provides, most people can do away with R's slightly higher cost in terms of time and leverage it's ease of use and integration with C++ for further optimization.

## VIII. FUTURE LEARNING

This paper touched various aspects of learning R for HPC, and has benchmarked Lower level languages and R. We have now have an excellent understanding of the actual core concepts of R in HPC, and we have several optimizations but we wanted to be brief, given that this paper was designed for beginners in R who want to start Scientific and Concurrent Computing. We could further delve into looking at and benchmarking more R-packages for concurrent and scientific computing and look at usign GPU's to accelerate R jobs with CUDA.

ACKNOWLEDGMENT

Computations for this research were performed on the Pennsylvania State University's Institute for CyberScience Advanced CyberInfrastructure (ICS-ACI). The Iris dataset was made available throughUCI Machine Learning Repository [http://archive.ics.uci.edu/ml]. Irvine, CA: University of California, School of Information and Computer Science.